\documentclass[fleqn,usenatbib]{mnras}

\usepackage{newtxtext,newtxmath}
\usepackage[T1]{fontenc}

\DeclareRobustCommand{\VAN}[3]{#2}
\let\VANthebibliography\thebibliography
\def\thebibliography{\DeclareRobustCommand{\VAN}[3]{##3}\VANthebibliography}

\usepackage{graphicx}
\usepackage{amsmath}	

\usepackage{wasysym}           
\usepackage{graphicx}
\usepackage{epstopdf}
\usepackage{mathrsfs}
\usepackage{anyfontsize}
\usepackage{natbib}
\usepackage{color}
\usepackage{lipsum}
\usepackage{diagbox}
\usepackage[thinc]{esdiff}
\usepackage[normalem]{ulem}

\title[Relativistic Settling Solutions]{Spherically Symmetric Accretion onto a Compact Object through a Standing Shock: The Effects of General Relativity in the Schwarzschild Geometry}

\author[Kundu \& Coughlin]{
Suman Kumar~Kundu$^{1}$\thanks{E-mail:skundu@syr.edu},
Eric R.~Coughlin$^{1}$\thanks{E-mail: ecoughli@syr.edu}
\\
$^{1}$Department of Physics, Syracuse University, Syracuse, NY 13244, USA}

\date{Accepted XXX. Received YYY; in original form ZZZ}

\pubyear{2022}

\begin{document}
\label{firstpage}
\pagerange{\pageref{firstpage}--\pageref{lastpage}}
\maketitle

\begin{abstract}
A core-collapse supernova is generated by the passage of a shockwave through the envelope of a massive star, where the shock wave is initially launched from the ``bounce'' of the neutron star formed during the collapse of the stellar core. Instead of successfully exploding the star, however, numerical investigations of core-collapse supernovae find that this shock tends to ``stall'' at small radii ($\lesssim$ 10 neutron star radii), with stellar material accreting onto the central object through the standing shock. 
Here, we present time-steady, adiabatic solutions for the density, pressure, and velocity of the shocked fluid that accretes onto the compact object through the stalled shock, and we include the effects of general relativity in the Schwarzschild metric. Similar to previous works that were carried out in the Newtonian limit, we find that the gas ``settles'' interior to the stalled shock; in the relativistic regime analyzed here, the velocity asymptotically approaches zero near the Schwarzschild radius. These solutions can represent accretion onto a material surface if the radius of the compact object is outside of its event horizon, such as a neutron star; we also discuss the possibility that these solutions can approximately represent the accretion of gas onto a newly formed black hole following a core-collapse event. Our findings and solutions are particularly relevant in weak and failed supernovae, where the shock is pushed to small radii and relativistic effects are large.
\end{abstract}

\begin{keywords}
hydrodynamics ---  methods: analytical  --- (stars:) supernovae: general --- shock waves
\end{keywords}
\section{Introduction}
 At the end of the life of a massive star, where it has exhausted its supply of nuclear fuel, the core collapses $\sim$ dynamically under its own self-gravity. The process of inverse-$\beta$ decay during the collapse removes electron pressure, further destabilizing the core and producing an abundance of neutrons; this de-leptonized, collapsing core is the proto-neutron star (PNS). 
 The PNS ``bounces'' due to the nucleon-nucleon interaction potential (that is, the equation of state of the nuclear material stiffens) and launches an outward-propagating shock wave. The shock dissociates heavy nuclei as it propagates outward, loses energy, and eventually stalls under the ram pressure of the infalling envelope \citep{Colgate66,Bethe79,Woosley86,Bethe85,Herant92,Herant94}. 
 
Some mechanism revives the shock and yields the powerful and luminous explosion that is the supernova. This ``supernova problem,'' being the stall and eventual revival (in successful explosions) of the shock wave launched by the bounce of the PNS, has eluded a conclusive theoretical explanation for decades. 
Two likely mechanisms for accelerating the stalled shock are 
the neutrino mechanism \citep{Bethe85,Bethe90,Murphy08,Ugliano12,Nakamura15,Chan18} and the standing accretion shock instability \citep{Blondin03,Blondin06,Blondin07,ohnishi08,Burrows12}. 
The neutrino mechanism, first laid out in the 1960s \citep{Colgate66,Arnett66} and revived in the mid-1980s by works such as \citet{Bethe85,Bruenn1985}, propose that some fraction of the neutrino flux radiated from the nascent neutron star is absorbed by the abundant free neutrons and protons in the post-shock layer; the energy and momentum deposited by the absorbed neutrinos then revives the stalled shock. The standing accretion shock instability arises from the fact that some of the large-angle (i.e., small spherical harmonic number $\ell$), oscillatory modes of the accretion shock can be dynamically unstable, the instability drives the shock outward leading to an asymmetric explosion \citep{Blondin03}.
Subsequent work (e.g., \citealt{Marek09}) has shown that hydrodynamic instabilities can aid the neutrino mechanism if not independently capable of driving the explosion.

Independent of the mechanism responsible for reviving the shock, it has been found to stall in many numerical calculations. Newtonian solutions describing the velocity, density, and pressure of the post-shock gas were found by \citealt{Lidov57,Chevalier89,Houck92}. In these solutions, the gas ``settles'' and the velocity of the fluid approaches zero asymptotically close to the origin for sufficiently small values of the adiabatic index (see the left panel of Figure \ref{fig:f_fa_Newtonian} below). 
\citet{Blondin03} obtained these solutions in the adiabatic limit (\citealt{Houck92} included the effects of cooling) and analyzed their response to angular perturbations.
These solutions and the numerical work of \citet{Blondin03} were in the Newtonian limit, and the gravitational field around the central compact object (PNS) was described as a Newtonian point mass.

However, for a neutron star with mass $M \,\,(\sim 1.4 M_\odot)$ and radius $R \,\, (\sim 10$ km), the gravitational radius is $GM/(Rc^2) \simeq 0.2$, and the free-fall speed of the shock radius at $R_{\rm sh} \simeq 100$ km is $\sqrt{2GM/R_{\rm sh}} \simeq 0.2\,c$. 
Relativistic effects therefore introduce order-unity corrections to the behavior of the gas within the shock and will non-trivially modify the Newtonian settling solutions.  
\cite{Michel72} expanded the classic work of \cite{Bondi1952} on spherically symmetric accretion by incorporating the effects of general relativity in the Schwarzschild metric (see also \citet{Blumenthal76} who expanded \citet{Michel72}’s work to non adiabatic equations of state). Bondi accretion (and its relativistic generalization) does not account for the existence of a shock\footnote{\citet{Blumenthal76} briefly discuss the possibility of a shock transition in the critical flow, which is different from the flow through an existing shock. It is also worth noting that those same authors used special relativistic shock jump conditions, which are inconsistent with the Schwarzschild background (see Equations \ref{eq:Jump1},\ref{eq:Jump2}, and \ref{eq:Jump3} below).}.  When the freely falling fluid passes through an existing strong shock, a substantial fraction of its kinetic energy is converted into internal energy -- an effect that cannot be considered a small perturbation on top of a pure freefall solution.
Subsequent work (e.g., \citealt{Fukue87,Chakrabarti89,Chakrabarti93,Gu03}) has also explored similar topics. However, the generalization of the \citet{Lidov57}, adiabatic settling solution {through an existing standing shock} to the Schwarzschild metric {incorporating the appropriate jump conditions} -- for which there are exact solutions (as we show below) -- does not appear to have been detailed in the literature.

Here, we present and analyze the relativistic generalization of adiabatic settling solutions for the post-standing accretion shock flow that were studied numerically in \citet{Blondin03}. 
In Section \ref{sec:fluid} we describe the model and write down the fluid equations. The ambient fluid -- assumed to be freely falling and effectively pressureless -- is analyzed in Section \ref{sec:ambient}, and we give the relativistic jump conditions in Section \ref{sec:jump}. Stationary solutions to the fluid equations that satisfy the relativistic jump conditions and are adiabatic are presented in Section \ref{sec:solutions}, where we also discuss the variation of the solutions with the shock radius (effectively the ambient velocity at the location of the shock), the variation with the adiabatic index and the behavior of the flow near the horizon. We present the physical interpretation of these solutions in Section \ref{sec:physical}, discuss the implications of our findings, and identify directions for future work in Section \ref{sec:sandc}.
\section{Equations}
\label{sec:relativistic}
\subsection{Metric}
\label{sec:metric}
We assume that there is a compact object (without spin) that dominates the gravitational field of the infalling fluid. With these assumptions, the metric describing the spacetime is given by the Schwarzschild metric: 
\begin{equation} 
\begin{split}
  ds^2&=g_{\mu\nu}dx^{\mu}dx^{\nu} \\
  &=-\left(1-\frac{2M}{r}\right)dt^2+\left(1-\frac{2M}{r}\right)^{-1} dr^2+r^2d\Omega^2, \label{metric} 
  \end{split}
\end{equation}
where $M$ is the mass of the compact object. We have adopted the Einstein summation convention (as we will throughout the remainder of the paper) so that repeated upper and lower indices imply summation, and we have let $G = c = 1$. 
\subsection{Fluid Equations}
\label{sec:fluid}
We let the accreting gas -- which has passed through the standing shock -- be a relativistic perfect fluid with total energy $e'$, pressure $p'$ and rest-mass density $\rho'$. For simplicity, we assume that the gas is adiabatic, with $e' = p'/(\gamma-1)$ and $\gamma$ the adiabatic index. With $U^{\mu}$ the four-velocity of the fluid, the energy-momentum tensor 
is \citep{anile_1990}
\begin{equation}
   \label{eq:E-M tensor}
    T^{\mu \nu}
    =\left(\rho'+\frac{\gamma}{\gamma-1}p'\right) U^\mu U^\nu+p'g^{\mu \nu}. 
\end{equation} 
Energy-momentum conservation is expressed as
\begin{equation} 
\label{eq:Conservation of E-M}
    \nabla_\mu T^{\mu\nu}=0
\end{equation}
where $\nabla_\mu$ is the covariant derivative. 
Conservation of mass (or particle number) is

\begin{equation} 
\label{eq:covcont}
    \nabla_\mu[\rho'U^\mu]=0,
\end{equation}
and from Equation \eqref{metric} we have the conservation of the norm of the four-velocity: 

\begin{equation} 
\label{eq:fourVel}
    U_\mu U^\mu=-1.
\end{equation}

It is useful to work with the time and space-like projections of Equation \eqref{eq:Conservation of E-M}  \citep{anile_1990,Coughlin19}. Taking the contraction of Equation \eqref{eq:Conservation of E-M} with $U_\nu$ yields

\begin{equation} 
\label{eq:covcoe}
    U^\mu \nabla_\mu e'= -(e'+p') \nabla_\mu U^\mu.
\end{equation}
We now introduce the projection tensor $\Pi_\nu^\beta=U^\beta U_\nu+g_\nu^\beta$, which projects the components of Equation \eqref{eq:Conservation of E-M} onto the 3-space orthogonal to $U^\mu$; contracting Equation \eqref{eq:Conservation of E-M} with the projection tensor then gives the momentum equations,

\begin{equation} 
\label{eq:cmom}
    (e'+p') U^\mu \nabla_\mu U^\nu +\Pi^{\mu \nu} \nabla_\mu p'=0.
\end{equation}
One can also re-express the energy equation in the following convenient form using the continuity Equation \eqref{eq:covcont},
\begin{equation} 
\label{eq:conentropy}
    U^\mu \nabla_\mu S'=0,
\end{equation}
where $S' = \ln\left(p'/\rho'^{\gamma}\right)$. This equation demonstrates that $S'$, which we interpret as the entropy of the gas, is a conserved Lagrangian quantity in adiabatic regions of the flow. Since we are assuming that the flow is spherically symmetric and irrotational, there are only two components of the four-velocity that are related by Equation \eqref{eq:fourVel}; we will refer to the radial component of the four-velocity by $U$.

\subsection{Ambient Fluid}
\label{sec:ambient}
 Pressure support is lost from the core, and a rarefaction wave travels through the overlying stellar envelope, which causes shells of material at successively larger radii to fall inward.  If the density and pressure of the ambient medium fall off as power-laws with distance from the core, one can show that there exists a self-similar solution to the fluid equations that describes the propagation of the rarefaction wave and the fluid interior to the wave \citep{Coughlin18}; the wave travels at the local sound speed, and the gas pressure of the fluid interior to the wave is much lower than the ram pressure. Therefore, the infalling gas can be treated effectively as pressureless. We denote the four-velocity of the ambient fluid by $U_{\rm a}=(U^{\rm t}_{\rm a}, U_{\rm a},0,0)$.
The radial momentum equation, Equation \eqref{eq:cmom} gives
\begin{equation}
    U_{\rm a} \diffp{U_{\rm a}}{r}=-\frac{M}{r^2} 
\end{equation}
Integrating the above equation and assuming that the gas is weakly bound, so the binding energy is $\sim 0$, then gives 

\begin{equation} \label{eq:ambientvellab}
 U_{\rm a}=-\sqrt{\frac{2M}{r}}.
\end{equation}
The time-steady solution to the continuity equation \eqref{eq:covcont} is then

\begin{equation}
 \rho'_{\rm a} = \bar{\rho}_{\rm a}\left(\frac{r}{R}\right)^{-3/2},
\end{equation}
where $R$ is the shock radius and $\bar{\rho}_{\rm a}$ is the density of the ambient gas at $r = R+\epsilon$ in the limit that $\epsilon \rightarrow 0$.

\subsection{Jump Conditions}
\label{sec:jump}
In the adiabatic limit the energy, mass, and momentum fluxes must be conserved across the shock, which give the strong-shock jump conditions -- assuming the ambient gas pressure is negligible -- in the lab frame (which equals the rest frame of the shock by assumption):

\begin{equation}
     \rho'_{\rm s} U_{\rm s} =\rho'_{\rm a} U_{\rm a} \label{eq:Jump1},
\end{equation}

\begin{equation}
         \left(\rho'_{\rm s}+\frac{\gamma}{\gamma-1}p'_{\rm s}\right) U_{\rm s} \sqrt{1-(U_{\rm a})^2+(U_{\rm s})^2}
    = \rho'_{\rm a}U_{\rm a} 
    \label{eq:Jump2}
\end{equation}

\begin{equation}
         \left(\rho'_{\rm s}+\frac{\gamma}{\gamma-1}p'_{\rm s}\right) (U_{\rm s})^2+  \left(1-(U_{\rm a }\right)^2) p'_{\rm s} =\rho'_{\rm a}U_{\rm a}^2
         \label{eq:Jump3}
\end{equation}
The subscript $``\rm{s}"$ indicates that these are the properties of the post-shock fluid at the shock radius. 

\begin{figure}
    \centering
    \includegraphics[width=0.475\textwidth]{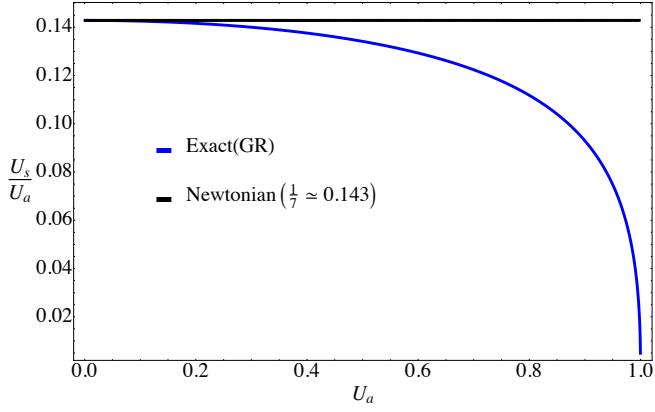}
    \caption{The ratio between the  post-shock fluid velocity to the ambient fluid velocity $({U_{\rm s}}/{U_{\rm a}})$ at the location of the shock as a function of the ambient fluid velocity, $U_{\rm a}$, in both the Newtonian and General Relativistic settings for $\gamma = 4/3$. Note that $U_{\rm a}$ is the freefall speed at the shock normalized by the speed of light, and hence the relativistic solution approaches the non-relativistic one when $U_{\rm a} \ll 1$; in the Newtonian limit, the entire problem is self-similar and $U_{\rm a}$ can be set to one without loss of generality. }
    \label{fig:stationaryjumpsol}
\end{figure}
Equations \eqref{eq:Jump1}, \eqref{eq:Jump2} and \eqref{eq:Jump3} can be combined into the following cubic \footnote{This equation reduces to the special relativistic jump conditions obtained in \citealt{Coughlin19} (Equation 23 therein with $U^{''}_2=U_{\rm s}$ and $U^{''}_a=U_{\rm a}$).} to be solved for $U_{\rm s}$:
\begin{equation} 
\label{eq:mycubic}
\begin{split}
     \gamma^2 U_{\rm s}^3+ (2 U_{\rm a} \gamma-U_{\rm a} \gamma^2)  U_{\rm s}^2+ \left(-1+U_{\rm a}^2+\gamma^2-U_{\rm a}^2 \gamma^2 \right)\\+\left((1-U_{\rm a}^2)U_{\rm a}+2 U_{\rm a} \gamma-2 U_{\rm a}^3 \gamma - U_{\rm a} \gamma^2 + \gamma^2 \right) = 0.
\end{split}
\end{equation}
Figure \ref{fig:stationaryjumpsol} shows the ratio $U_{\rm s}/U_{\rm a}$ resulting from Equation \eqref{eq:mycubic} 
as a function of the ambient four-velocity. We see that the ratio is nearly equal to its Newtonian value $(\sim 0.14)$ for $U_{\rm a} \lesssim 0.2$, but as the ambient velocity becomes more relativistic, the ratio deviates significantly from the Newtonian value.

\subsection{Bernoulli, mass, and entropy conservation equations}
\label{sec:dynamics}
Combining Equations \eqref{eq:covcont} and \eqref{eq:covcoe} yields the conservation of the radial energy flux,
\begin{equation}
\label{eq:postshockenergy}
    \left(1+\frac{\gamma}{\gamma-1} \frac{p}{\rho'}'\right)^2 \left(1-\frac{2M}{r}+U^2\right)=\Dot{E}
\end{equation}
We use the continuity Equation \eqref{eq:covcont} to express the density of the post-shock fluid in terms of the four-velocity as
\begin{equation}
  r^2\rho'U=\Dot{M}, \label{mdot}
\end{equation}
while entropy conservation gives

\begin{equation}
    p'=K(\rho')^\gamma. \label{entropy}
\end{equation}
With these results, we can write Equation \eqref{eq:postshockenergy} as

\begin{equation}
\label{eq:postshockenergy2}
    \left(1+\frac{\gamma}{\gamma-1} K\bigg(\frac{\Dot{ M}}{r^2 U}\bigg)^{\gamma-1}\right)^2 \left(1-\frac{2 M}{r}+U^2\right)=\Dot{E}.
\end{equation}
Equation \eqref{eq:postshockenergy2} is the relativistic generalization of the Bernoulli equation, to which it manifestly reduces in the limit that the velocity is small (see Equation 4 in \citealt{Blondin03}).

\section{General Relativistic Accreting Solutions}
\label{sec:solutions}
\subsection{Impact of varying the ambient fluid velocity}

\begin{figure*}
    \centering
    \includegraphics[width=0.505\textwidth]{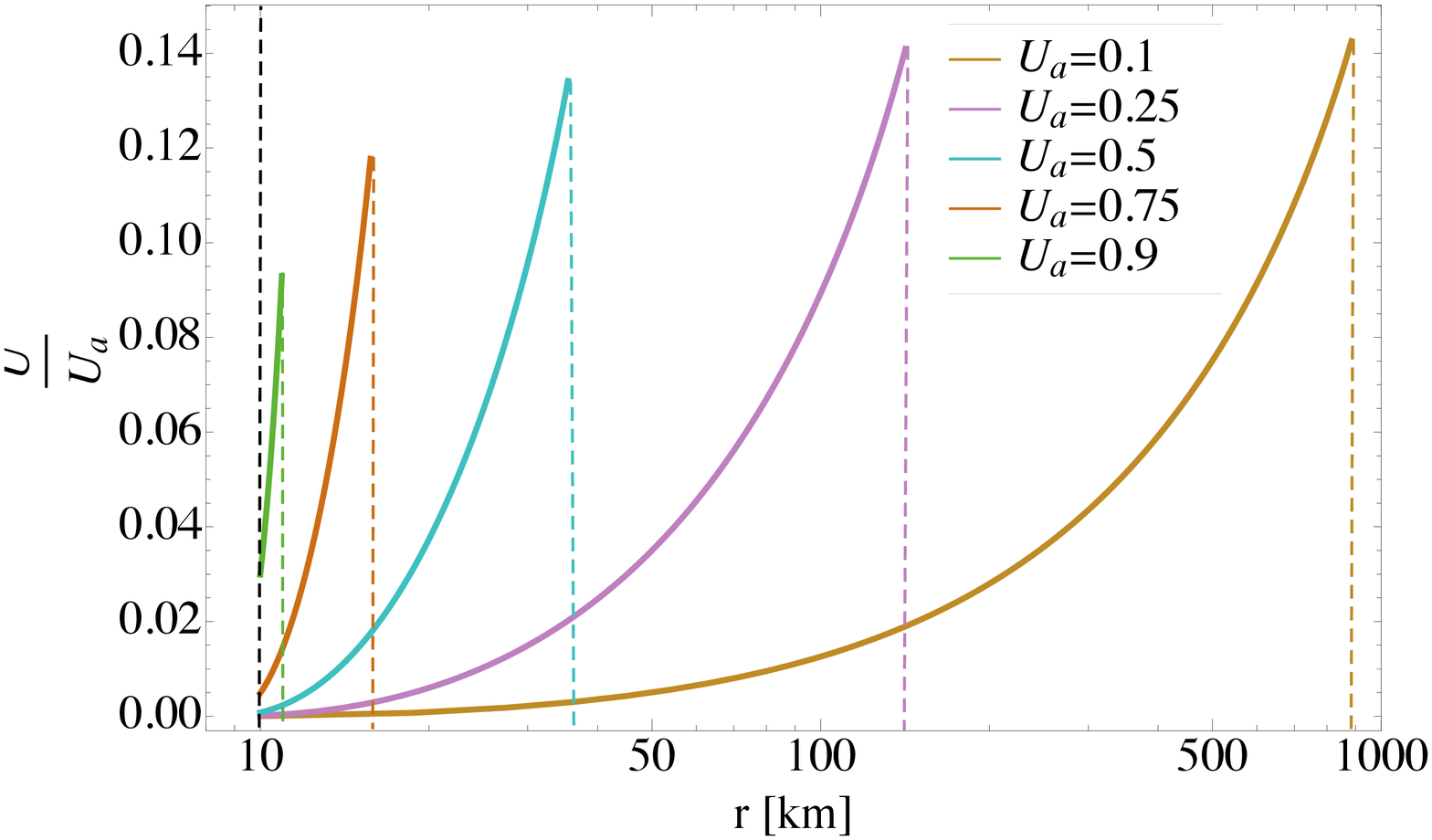}
    \includegraphics[width=0.475\textwidth]{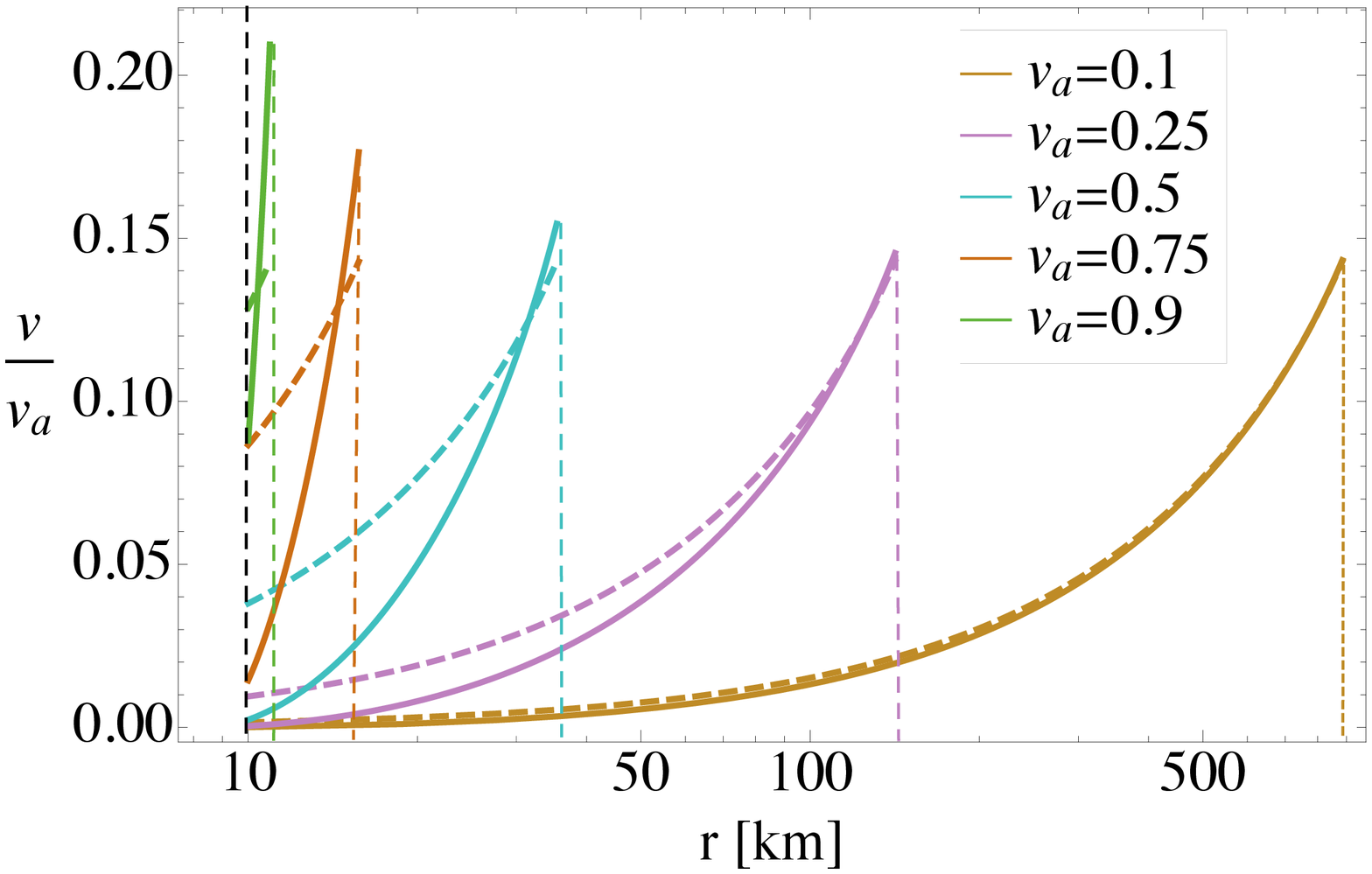}
    \newline
    \newline
     \includegraphics[width=0.49\textwidth]{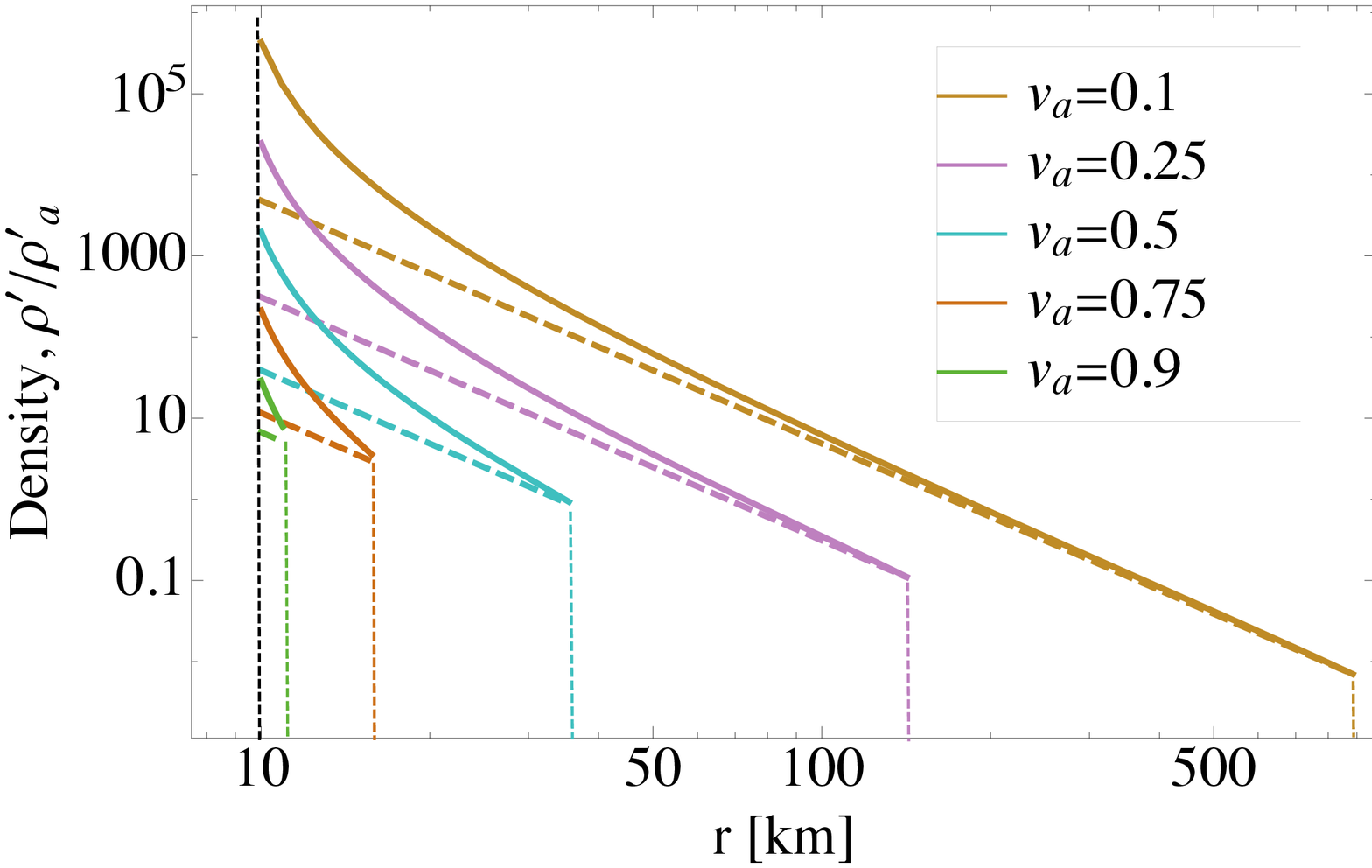}
     \includegraphics[width=0.49\textwidth]{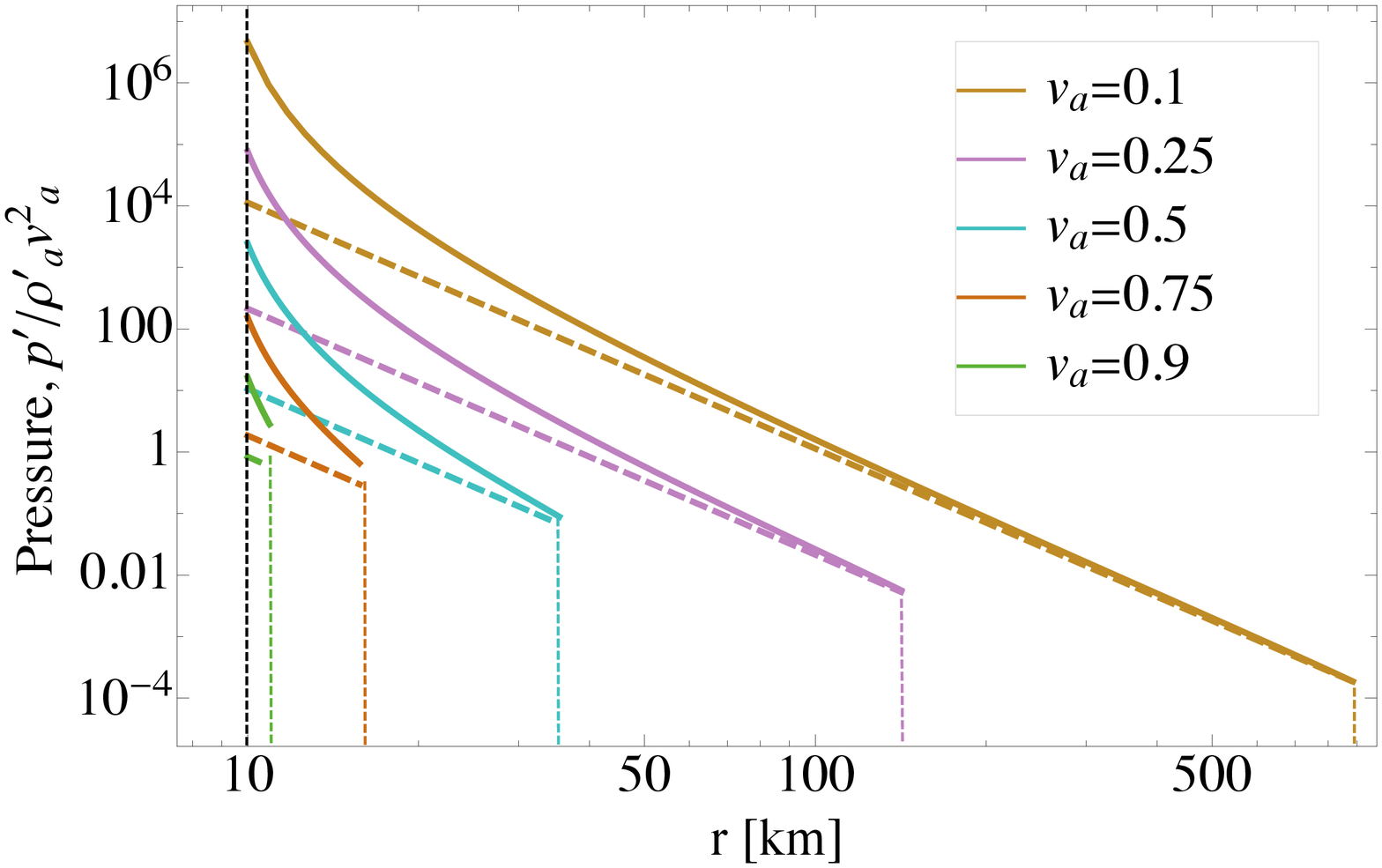}
     \caption{The normalized fluid velocity (the radial component of four-velocity in the top-left and the three-velocity as seen by local static observer in the top-right), comoving density (bottom-left), and comoving pressure (bottom-right) for the ambient speeds (at the location of the shock) shown in the legends as functions of radius (in km). Here we set the neutron star mass to be $3\,M_{\odot}$ to convert to physical units. The black dashed line indicates the surface of the neutron star whereas the colored dashed lines indicate the location of the stalled shock appropriate to the respective ambient velocity. Newtonian limits are given by the dashed curves in each panel.}
    \label{fig:grsol}
\end{figure*}

\begin{figure}
    \centering
    \includegraphics[width=0.48\textwidth]{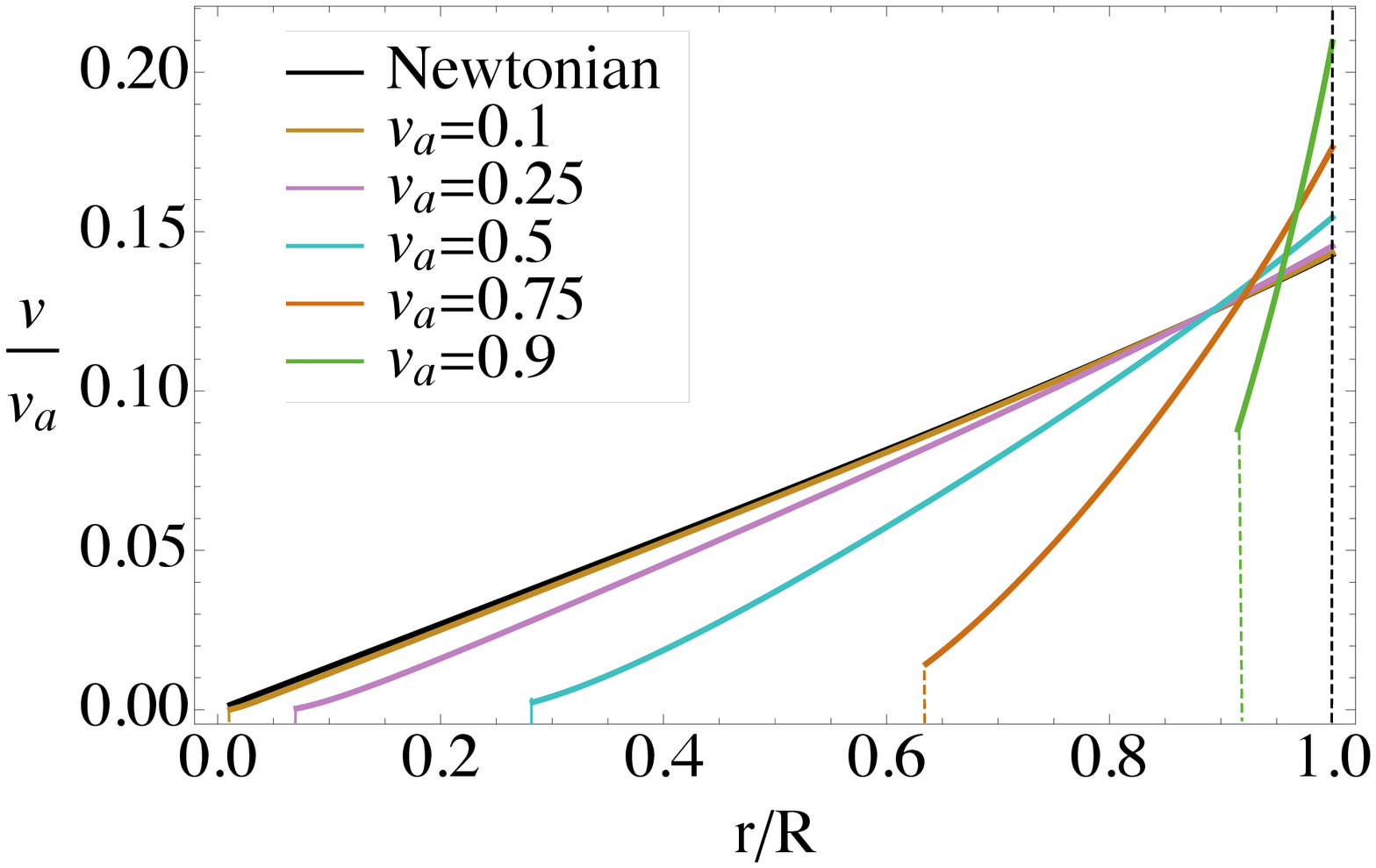}
     \caption{The normalized three-velocity as seen by a static and locally flat observer for the ambient speeds (at the location of the shock) shown in the legend as functions of normalized radial coordinate by the shock radius ($r/R$); the black dashed line at $r/R=1$ indicates the fixed location of the shock in this coordinate and the colored dashed lines indicate the location of the neutron star surface, assumed to be at 10 km. The Newtonian solution, which is self-similar (i.e., it does not depend on $v_{\rm a}$), is shown by the solid black curve.}
    \label{fig:grsol2}
\end{figure}

Here, we discuss the effects of varying the velocity of the ambient fluid, $U_{\rm a}$, while keeping $\gamma$ fixed at ${4}/{3}$. Given the fluid velocity $U_{\rm a}$, we can calculate the entropy $K=p'/(\rho')^{4/3}$, the radial energy flux $\Dot{E}$ and the radial mass flux $\Dot{M}$ by using the jump conditions and we can then solve Equation \eqref{eq:postshockenergy2} numerically for the post-shock fluid velocity $U(r)$. We can then calculate the fluid three-velocity as seen by an observer who is stationary with respect to the compact object and who employs locally flat coordinates (this coordinate frame will be represented with `hats') as
\begin{equation}
    v^{\hat{r}}=\frac{U^{\hat r}}{U^{\hat{t}}}=\frac{U}{U^t}\frac{1}{1-2M/r} =\frac{U}{(1-2M/r+U^2)^{1/2}}
    \label{eq:3vel},
\end{equation}
where in the last equality we used Equation (\ref{eq:fourVel}). 
For a neutron star of mass $3 \, \rm{M_\odot}$ and radius\footnote{We use 10 km for the neutron star radius for concreteness and simplicity, though \citet{lattimer16} and \citet{haensel99} find that causality arguments require that the neutron star radius satisfy $R \gtrsim 2.823GM/c^2$, which is closer to 12.5 km for $M = 3M_{\odot}$. If we used 12.5 km, the maximum speed able to be achieved by the infalling fluid -- obtained when the shock radius is comparable to the neutron star radius -- would be $\sim 84\%$ c instead of $\sim 90\%$ c, as shown in Figure \ref{fig:grsol}.} $10 \, \rm{km}$, the solutions for the post-shock fluid four-velocity, three-velocity, density and pressure are presented respectively in the top-left, top-right, bottom-left and bottom-right panels in Figure \ref{fig:grsol}; the ambient four-velocity -- which implicitly establishes the physical location of the shock since the mass of the neutron star is set -- for each curve is shown in the legend.  
In Figure \ref{fig:grsol}, the radial coordinate is in $\rm{km}$, while in Figure \ref{fig:grsol2} it is normalized by the shock radius. In Figure \ref{fig:grsol}, the location of the neutron star surface is shown with a black dashed line, and the location of the shock, appropriate to the specific ambient velocity, is shown by the respective colored dashed line. In Figure \ref{fig:grsol2} we show the three-velocity as a function of $r/R$, in which case the shock is always at $r/R=1$, represented by a black dashed line, while the neutron star surface now takes different $r/R$ values and is shown with colored dashed lines. Figure \ref{fig:grsol2} demonstrates that the relativistic solutions are not self-similar -- each curve displays qualitatively different behavior as a function of the ambient velocity, whereas the Newtonian limit (shown by the black curve) is independent of this quantity once it is scaled out of the solution. Both of these figures show that the relativistic solutions approximately equal the Newtonian solution in the limit that the ambient speed is non-relativistic, which is not surprising. However, substantial deviations arise when the ambient speed reaches substantial fractions of the speed of light, and this is especially true deep in the interior of the flow where relativistic gravity is yet more important.

\subsection{Impact of varying the adiabatic index}
In addition to the magnitude of the infall velocity (relative to the speed of light), the solutions for the post-shock fluid variables depend on the adiabatic index of the gas. The $\gamma={4}/{3}$ adiabatic fluid is likely a reasonable approximation of the radiation-pressure dominated post-shock fluid accreting onto a neutron star \citep{Chevalier89, Houck92} over the expected temperature and density ranges \citep{Schinder87}. Although in our model we neglected non-ideal effects (e.g., neutrino cooling), these can be roughly captured by using a softer equation of state, i.e., reducing the value of the adiabatic index. While it is difficult to see how they are physically relevant, we also analyze the $\gamma={5}/{3}$ monoatomic ideal gas and one or two cases with even higher adiabatic indices, primarily to compare with \citet{Blondin03}.

\begin{figure*}
    \centering
    \includegraphics[width=0.475\textwidth]{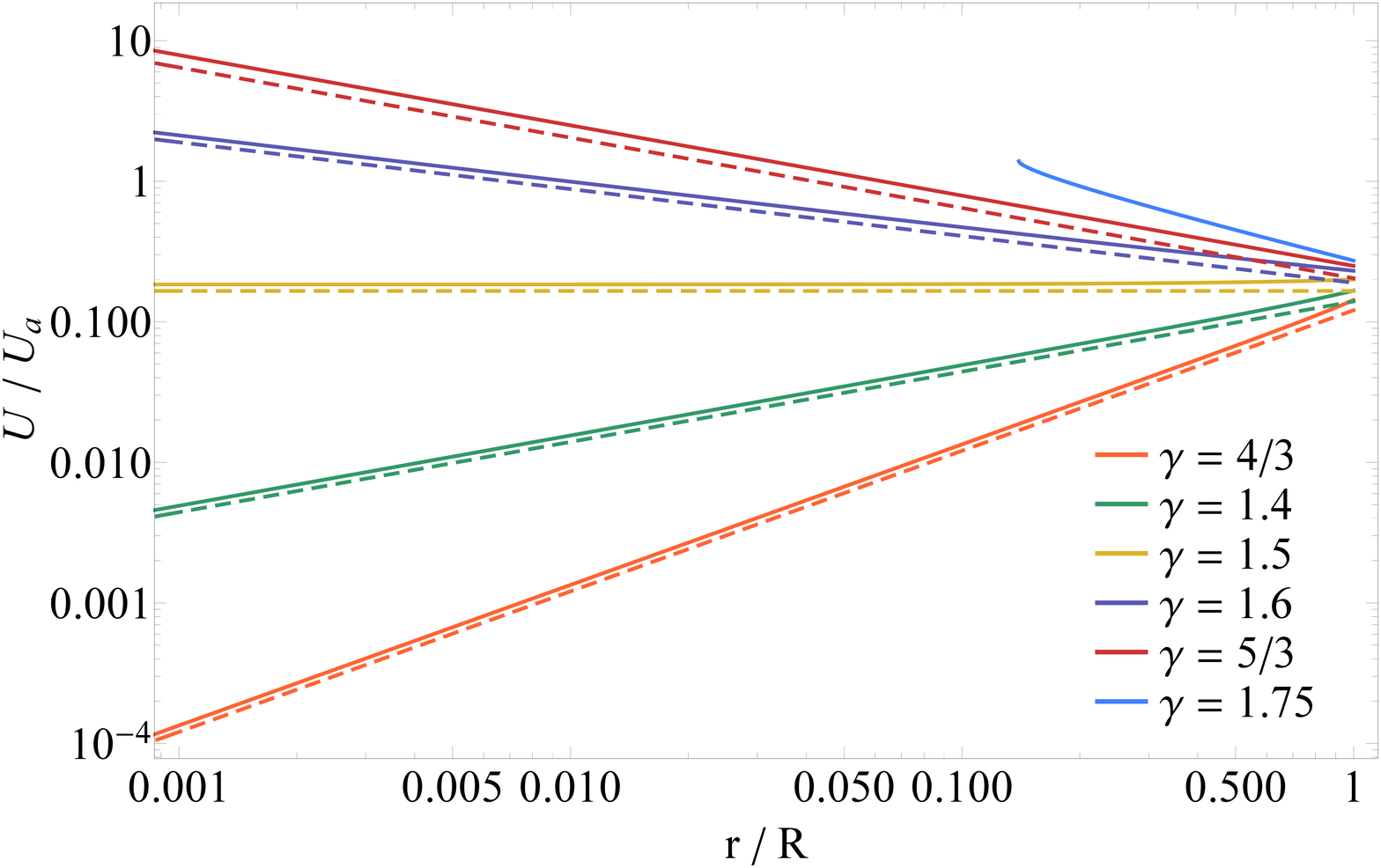}
     \includegraphics[width=0.495\textwidth]{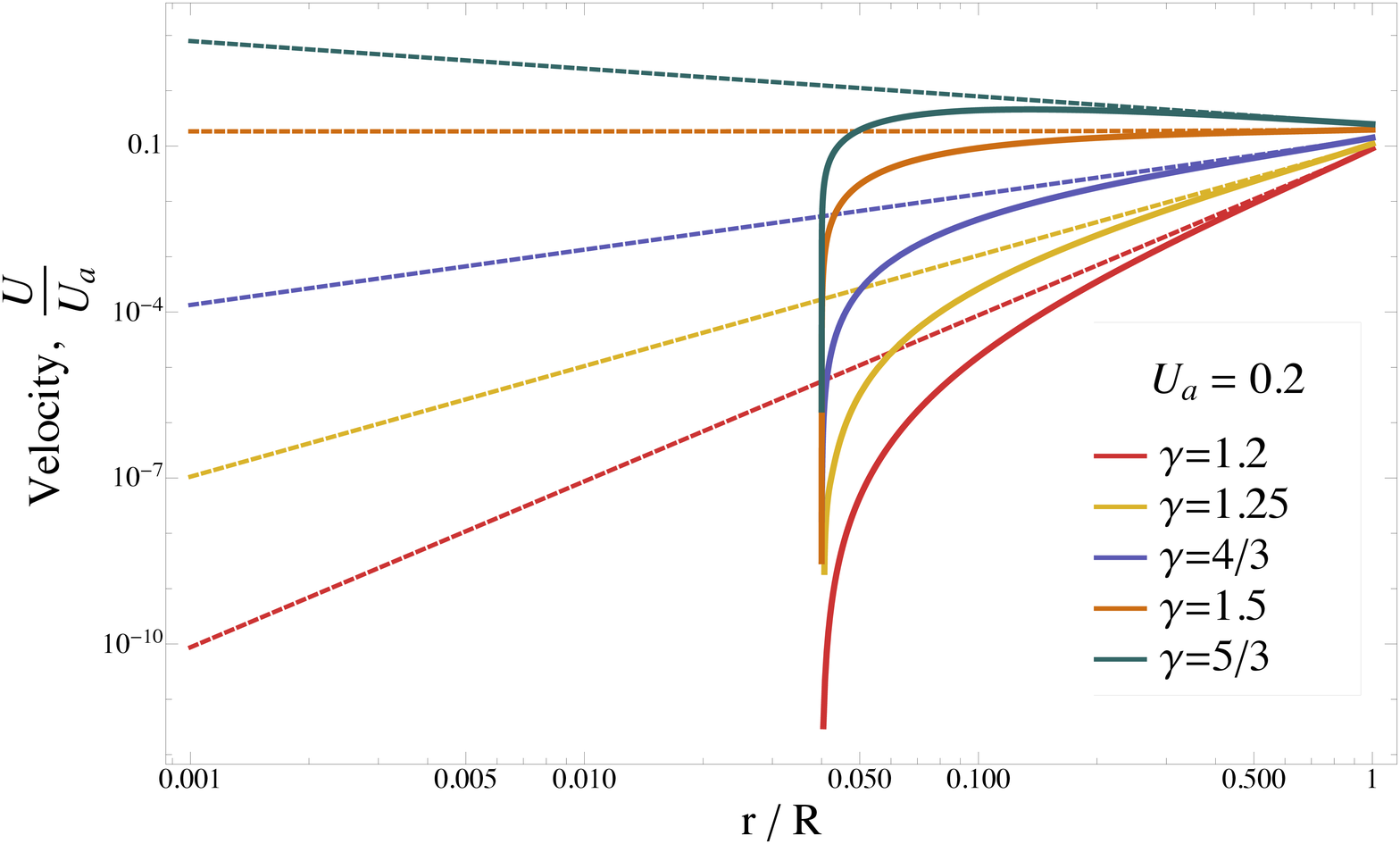}
    \caption{Left: The Newtonian solution for the fluid velocity as a function of radius normalized by the shock radius for the adiabatic indices in the legend; the solid curves show the numerical solution from the Bernoulli equation in the Newtonian limit, while the dashed curves give the analytic, asymptotic scalings, arbitrarily offset by a factor of 0.9 so that they can be differentiated from the exact solutions when $r/R\ll 1$. Right: The fluid velocity derived from the relativistic Bernoulli equation (solid curves) and the Newtonian Bernoulli equation (dashed curves) for an ambient velocity of $U_{\rm a} = 0.2$ and the adiabatic indices in the legend. The two solutions agree well near the shock front but disagree strongly near the horizon, which occurs at $r/R = 0.04$ in this case.}
    \label{fig:f_fa_Newtonian}
\end{figure*}

The left panel of Figure \ref{fig:f_fa_Newtonian} illustrates the absolute value of the velocity as a function of distance behind the shock in the Newtonian limit for the adiabatic indices in the legend. The solid curves show the solution to the Bernoulli equation in the non-relativistic limit (i.e., Equation \ref{eq:postshockenergy2} when the rest-mass energy far outweighs the internal energy and $GM/(rc^2) \ll 1$; see also Equation 4 in \citealt{Blondin03}), while the dashed curves give the asymptotic scaling near the origin\footnote{ We have arbitrarily scaled the analytic, asymptotic solutions by a factor of 0.9 so that they can be distinguished from the exact solutions in this figure.}

\begin{equation}
    \frac{U}{U_{\rm a}} = \left(\frac{4\gamma}{\gamma^2-1}\left(\frac{\gamma-1}{\gamma+1}\right)^{\gamma}\right)^{\frac{1}{\gamma-1}}\left(\frac{r}{R}\right)^{\frac{1}{\gamma-1}-2}. 
    \label{fasymp}
\end{equation}
This scaling can be derived from Equation \eqref{eq:postshockenergy2} in the Newtonian limit and assuming that the internal energy far outweighs the kinetic energy near the origin, which is a valid assumption when $\gamma < 5/3$. 

Equation \eqref{fasymp} also shows, and the left panel of Figure \ref{fig:f_fa_Newtonian} verifies that the velocity declines in absolute magnitude as one approaches the origin for $\gamma < 1.5$, while the gas accelerates (in terms of the magnitude of the velocity) for $\gamma > 1.5$; we believe that the value of $\gamma_{\rm c} = 1.522$ quoted in \citet{Blondin03} that differentiates between these two limits is in error -- it is only for $\gamma \equiv 1.5$ that the velocity approaches a finite value (i.e., neither zero nor infinite) near the origin. For $\gamma = 5/3$ the velocity satisfies $U/U_{\rm a} = 0.25 (r/R)^{-1/2}$ and scales exactly with the freefall speed, as also found in \citet{Blondin03}. For $\gamma > 5/3$, Equation \eqref{fasymp} predicts that the fluid speed increases in a manner that exceeds the freefall scaling as the gas approaches the origin. However, this is not consistent with the assumption that the kinetic energy remains sub-dominant to the thermal energy, upon which the assumption Equation \eqref{fasymp} is based. Instead, the blue curve in the left panel of Figure \ref{fig:f_fa_Newtonian} demonstrates that this super-freefall acceleration is only maintained for a finite distance beneath the shock, and that the solutions terminate at a sonic point (i.e., the derivative of the velocity diverges at radius of $r/R \simeq 0.14$ for $\gamma = 1.75$, and solutions do not exist for radii smaller than this value). While this behavior is interesting from an academic standpoint, it is difficult to see how such stiff equations of state could be realized in nature, and hence we do not consider these solutions further here.

In the right panel of Figure \ref{fig:f_fa_Newtonian}, we present the solution to the relativistic Bernoulli equation (Equation \ref{eq:postshockenergy2}) by solid curves, where the adiabatic index appropriate to each curve is given in the legend. We set the ambient fluid velocity to $0.2c$ for all solutions as a fiducial value. The dashed curves represent the corresponding Newtonian solutions. We see that the fluid ``settles" at the event horizon, instead of conforming (approximately) to power-laws near the origin, as is the case in the Newtonian approximation. The Newtonian solutions display qualitatively different behavior above and below $\gamma = 1.5$, as illustrated in the left panel of Figure 3, with the gas decelerating (accelerating) for $\gamma <1.5$ ($\gamma > 1.5$); on the other hand, the relativistic solutions all decelerate and settle as we approach the horizon, even though they closely match the Newtonian solutions near the shock. The asymptotic scaling of these solutions near the horizon is presented in the next subsection.

\subsection{Asymptotic behavior of the general relativistic solutions near the horizon}
\label{sec:asymp}
The neutron star surface -- where the fluid must physically stop -- is always outside of the horizon, but it is instructive to discuss the extreme limit where the neutron star surface approaches the horizon.
The fluid variables as measured by an observer in the comoving and locally flat frame either approach zero (velocity) or diverge (density and pressure) near the Schwarzschild radius, and it is straightforward to show from Equations \eqref{mdot}, \eqref{entropy}, and \eqref{eq:postshockenergy2} that the rates at which they do so are

\begin{equation}
    U \propto \left(1-\frac{2M}{r}\right)^{\frac{1}{2(\gamma-1)}},
\end{equation}

\begin{equation}
    \rho' \propto \left(1-\frac{2M}{r}\right)^{-\frac{1}{2(\gamma-1)}},
\end{equation}
and 

\begin{equation}
    p' \propto \left(1-\frac{2M}{r}\right)^{-\frac{\gamma}{2(\gamma-1)}}.
\end{equation}
Thus, in the general relativistic solutions presented here, the fluid velocity ``settles'' as in the Newtonian regime. However, instead of settling near the origin, the fluid velocity approaches zero near the event horizon. Similarly, the density and pressure diverge at small radii, but instead of following $\rho \propto r^{-\frac{5-3\gamma}{\gamma-1}}$ and $p \propto r^{-\frac{(5-3\gamma)\gamma}{\gamma-1}}$ as they do in the non-relativistic regime, they diverge as simple (and smaller) powers of $1-2M/r$.

Due to the presence of the strong shock, the fluid loses most of of its kinetic energy (gains equivalent thermal energy) and hence inevitably comes to rest interior to the shock. As it approaches the horizon it sees an unbounded gravitational field and can therefore only decelerate by developing an infinite pressure gradient as the event horizon is approached. 
We turn to the implications of these findings in the next section and our corresponding physical interpretations. 

\section{Physical Interpretation}
\label{sec:physical}
The settling solutions obtained by \citet{Lidov57}, \citet{Chevalier89} and \citet{Blondin03} in the Newtonian limit possess the feature that the velocity settles, or approaches zero, near the origin, which coincides with the location of the compact object in the Newtonian point-mass limit. This feature seems to imply that the Newtonian settling solutions describe the accretion of material onto the surface of an object and that these solutions apply to an accreting neutron star during the initial, stalled phase of the bounce shock, or at later times when weakly bound material falls back from the expulsion of the envelope in a successful explosion (e.g., \citealt{Chevalier89}). The pressure and density of the gas also rise dramatically near the origin in the Newtonian settling solutions, which suggests that the pressure of the gas as it is brought to a halt at the surface of the neutron star provides the force that decelerates the flow.

However, while the velocity of the Newtonian solution approaches zero near the origin, the \emph{mass flux} remains finite, as it must by virtue of the time-steady nature of the accretion through the shock. Similarly, the energy flux is conserved throughout the domain, and is given by the kinetic energy flux across the shock. Both of these quantities must effectively vanish at the surface of the neutron star where the velocity is (again, effectively) zero and the density and pressure retain finite values.
Thus, if the solutions here are to describe accretion onto a neutron star (or any object with a ``surface'' at which the equation of state stiffens substantially), then the mass flux should not contribute substantially to the mass of the compact object, and there should be a mechanism for removing the incident energy flux. In core-collapse supernovae, both of these conditions are approximately upheld over the freefall time from the shock (i.e., conditions are roughly time-steady): the mass accumulated near the surface is small compared to the mass of the star, and neutrino losses from a very thin layer near the surface negate the incoming energy flux. Thus, the relativistic solutions here should apply to accretion onto a neutron star and become particularly relevant for scenarios in which the shock is pushed to small radii (see the additional discussion in Section \ref{sec:sandc} below).

In failed supernovae, the continued accretion onto the neutron star eventually pushes it over the TOV-limit, and the star collapses dynamically to a black hole \citep{sumiyoshi07, sumiyoshi08, Connor11, Ugliano12, sukhbold16, kuroda22}. When the star collapses, an inward velocity must develop near the surface, and the pressure gradient must fall below the value necessary to retain hydrostatic balance. Thus, the solutions analyzed here cannot describe the entirety of the flow structure near the horizon when the black hole forms, which is also not expected because of the highly time-dependent nature of the interior of the flow as the core starts to collapse (i.e., the assumption of steady flow is clearly violated at this point). It is then possible that the shock is not supported by the reduced pressure in the interior and is correspondingly swallowed by the black hole.

On the other hand, it seems possible that the adiabatic increase in the pressure of the gas as it nears the horizon (as evidenced from Figure \ref{fig:grsol} above) is sufficient to support the shock for at least a finite time, even though the divergence of the pressure is not physical (the true location of the horizon must increase to accommodate the large increase in the density predicted by our solutions, which would result in non-divergent values of fluid variables at the horizon). One could analyze the effects of imposing a more negative velocity at an inner radius (mimicking the effects of a true horizon) on the flow by using linear perturbation analysis, similar to what was done in \citet{Blondin03} (though they used a smaller velocity, which caused the outward motion of the shock). The fading mass supply from the overlying envelope as less-dense regions of the progenitor are swallowed also implies that the ram pressure should be less capable of stifling the shock, which could allow the shock to remain quasi-stationary even in the absence of the strong pressure gradient in the interior that is predicted by the solutions here. 

For these solutions to arise physically (i.e., in a setting in which the assumptions made here are relaxed), a steady-state must be reached by the flow. In general, this will take on the order of a sound-crossing time from the shock, which for a shock radius of $R = 150$ km and neutron star mass of $M = 2M_{\odot}$ is $\sim R^{3/2}/\sqrt{GM} \sim 3.5$ ms. If the neutron star collapses to a black hole on a timescale that is shorter than this, then these solutions will not be realized and the flow will be better approximated by pure freefall (e.g., Figure 7 of \citealt{kuroda22}).

\section{Summary and Conclusions}
In this paper we analyzed the adiabatic accretion of gas through a stalled shock by modeling the gravitational presence of the compact body with the Schwarzschild geometry. 
The strong-gravitational effects (i.e., the inclusion of general relativity) is especially important in the case of weak or failed supernovae, where the shock wave launched by the bounce of the proto-neutron star could stall at much smaller, more relativistic radii, and during which a black hole forms. The black hole can continue to accrete at least as long as there remains supply from the host star, and we suggest that the solutions outlined here should also apply in this phase, i.e., while the black hole accretes through the stalled shock.

 For a neutron star mass of $1.4 M_{\odot}$ and a shock radius of $100 \text{ km}$, the freefall speed at the shock satisfies $\frac{v}{c} \sim 0.2$ and the differences between the relativistic and Newtonian solutions are at the level of $\sim 10$ \% (see the red curves in Figure 2). However, for failed supernovae (in which the shock is not revived), the shock can be pushed to smaller radii and the mass of the neutron star increases to near the TOV limit (maximum mass limit for neuron stars $\lesssim 3 M_{\odot}$ for most equations of state). For example, \citet {Connor11} find that the shock can stall at radii as small as $~20\text{ km}$; with a neutron star mass of $3 M_{\odot}$, the freefall speed at this radius satisfies $\frac{v}{c} \simeq 0.66$, and relativistic effects become much more important (see the green and orange curves in Figure 2). Once the star exceeds the TOV limit, the mass increases to even larger values, while the shock can be compressed to even smaller radii, which necessitates the usage of our solution over the Newtonian one. 

Relativistic effects will change the stability criteria and the corresponding growth of the standing accretion shock instability (SASI) (e.g., \citealt{Blondin03, Foglizzo_2007, Fernandez09a,Fernandez09b}), and these effects can be important because the growth rate of the SASI has been found to be small (i.e., the e-folding time of the instability is many freefall times). 
For example, \citet{Blondin07} find from their one- and two-dimensional simulations that the growth rate of SASI is between $\sim (0.1 - 0.22)$ in units of the freefall time (i.e., the instability grows as $\sim e^{(0.1-0.22)\tau}$, where $\tau$ is time relative to the freefall time; see their Figure 2). Their results agree well with those obtained analytically by \citet{Houck92}. \citet{Foglizzo_2007} and \citet{Fernandez14} too find similar, small growth rates. Small changes in the background state can therefore change the stability of the system, and the solutions here present one such instance in which small changes arise from physical (relativistic) considerations.

Our solutions also have implications for gravitational wave signals from core-collapse supernovae. For example, \citet{Morozova18} discuss the possibility that material that falls through the shock can impact the nascent neutron star and generate detectable gravitational wave signatures. Relativistic effects will modify the properties of the fluid as it rains down onto the neutron star surface, which could lead to pronounced differences in the gravitational wave signal owing to the fact that the neutron star radius is only marginally greater than the Schwarzschild radius. 

In this work we assumed spherically symmetric, irrotational flow. If the progenitor does not contain a large reservoir of bulk angular momentum (which is likely, because stellar winds will carry away a substantial amount of angular momentum over the lifetime of the star), then angular momentum is important during the late stages of infall; this is because it is only in the outer envelope of the progenitor where random, convective motions can yield a specific angular momentum that exceeds the innermost stable circular oribit (ISCO) value \citep{Quataert19}.
If the progenitor does have significant net angular momentum, the gas can circularize outside of the ISCO and rotation can be dynamically important. The disc-like structure that forms in this case is likely optically and geometrically thick owing to the extremely high accretion rates, and we can integrate the equations over the scale height of the disc to obtain height-averaged Euler equations. The resulting disc solutions would be analogous to the advection-dominated accretion flow solutions of \citealt{Narayan94} and the adiabatic inflow-outflow solutions of \citealt{Blandford04}, but with the added constraint of satisfying the boundary conditions at a shock. We defer further investigation of this possibility to future work.

We assumed that the post-shock fluid is adiabatic and that the time-steady nature of the flow then ensures that the gas is isentropic. The gas is expected to be nearly isentropic because the fluid in the gain region is convectively unstable \citep{Bruenn1985, Burrows87, Burrows1995}.  \citet{Chevalier89} justifies the adiabatic assumption by noting that neutrino losses become important in a very thin layer near the surface of the accreting body, and thus the post-shock flow is effectively adiabatic over most of its volume.  \cite{Muller20} notes the neutrino cooling is not likely to result in a significant entropy gradient across the flow in the post-shock region. Furthermore, in his $3D$ simulations, he finds that mixing at the onset of convection and SASI largely reduce any entropy gradient. Nevertheless, some simulations in lower dimensions $(2D)$ do find a more substantial entropy gradient, e.g., \citet{Muller10}. Interestingly, however, the numerical solutions of \citet{Muller10} appear to agree fairly well with the analytic solutions presented here. For example, in Figure $8$ of \cite{Muller10}, the density is shown to increase by $\sim 6$ orders of magnitude going from the location of the shock at $\sim 200 \rm{km}$ to the origin, which agrees with what we find analytically (see the purple curve in the bottom left panel of Figure \ref{fig:grsol}). The neutrino cooling in the thin layer acts effectively as a global sink on the energy, which balances the influx of energy from the material falling through the shock and allows the system to reach a steady state (i.e., the shock stalls). \citet{Chevalier89} also shows the ratio of thermal pressure to radiation pressure is
\begin{equation}
    \frac{p_{\rm th}}{p_{\rm rad}} \approx 0.02 \left(\frac{\dot{M}}{M_{\odot}}\right)^{0.4}
\end{equation}
for $\dot{M}=(0.01-100) \, M_{\odot}$ yr$^{-1}$ \citep{Schinder87,Chevalier89} reinforcing the adiabatic claim further. 

The time-steady nature of the solutions presented here requires specific ambient density and velocity profiles. More realistically, the mass infall rate will decline with time and the density profile of the accreting gas will change nontrivially as a consequence of the shells of nuclear ash in the progenitor star. The reduction in the mass supply rate and the ram pressure of the envelope then likely results in the outward motion of the stalled shock. We will analyze the consequences of a time-varying infall rate through a perturbative approach in future work. 
\label{sec:sandc}

\section*{Data Availability Statement}
Code to reproduce the results in this paper is available upon reasonable request to the corresponding author.
\section*{Acknowledgements}
We thank the anonymous referee for useful and constructive comments that improved the manuscript. SKK and ERC acknowledge support from the National Science Foundation through grant AST-2006684, and ERC acknowledges additional support from the Oakridge Associated Universities through a Ralph E.~Powe Junior Faculty Enhancement Award.
\bibliographystyle{mnras}
\bibliography{grsasi.bib}

\label{lastpage}
\end{document}